\documentclass[twocolumn,aps,prl,superscriptaddress,showpacs]{revtex4}
\usepackage{graphicx}

\begin{document}

\title{Non-Guassian normal diffusion induced by delocalization}
\author{Jianjin Wang}
\affiliation{Department of Physics and Institute of Theoretical Physics and Astrophysics,
Xiamen University, Xiamen 361005, Fujian, China}
\author{Yong Zhang}
\affiliation{Department of Physics and Institute of Theoretical Physics and Astrophysics,
Xiamen University, Xiamen 361005, Fujian, China}
\author{Hong Zhao}
\email{zhaoh@xmu.edu.cn}
\affiliation{Department of Physics and Institute of Theoretical Physics and Astrophysics,
Xiamen University, Xiamen 361005, Fujian, China}
\affiliation{Collaborative Innovation Center of Chemistry for Energy Materials, Xiamen
University, Xiamen 361005, Fujian, China}
\date{\today }

\begin{abstract}
Non-Gaussian normal diffusion, i.e., the probability distribution function
(PDF) is non-Gaussian but the mean squared displacement (MSD) depends on
time linearly, has been observed in particle motions. Here we show by
numerical simulations that this phenomenon may manifest in energy diffusion
along a lattice at a non-zero, finite temperature. The models we study are
one-dimensional disordered lattices with on-site potential. We find that the
energy density fluctuations are spatially localized if the nonlinear
interaction is suppressed, but may relax with a non-Gaussian PDF and a
linear time-dependent MSD when the nonlinear interaction is turned on. Our
analysis suggests that the mechanism lies in the delocalization properties
of the localized modes.
\end{abstract}

\maketitle


The non-Gaussian normal diffusion phenomenon was first evidenced
experimentally in supercooled liquids~\cite{Weeks, Kegel}. Later on it was
also observed in the diffusive process of polystyrene beads on the surface
of a lipid bilayer tube, of beads in an entangled solution of actin
filaments~\cite{Bwang}, in soft matter systems~\cite{Leptos, Kurtuldu, Juan}%
, on polymer thin films~\cite{Bhatt}, and in simulations of a
two-dimensional system of discs~\cite{Kim}. It refers to the phenomenon that
the probability distribution function (PDF) is non-Gaussian but the mean
squared displacement (MSD) increases as time linearly. This phenomenon was
believed to be common for particle diffusion in spatial and/or temporal
heterogeneous medium, particularly in a variety of physicochemical and
socioeconomic systems~\cite{Bwang}. There are two conjectures about the
mechanism. One assumes that a particle may experience different diffusion
processes with different diffusion coefficients; though in each individual
process it follows the normal Brownian motion, the composition of the
individual Gaussian PDFs may result in an overal non-Gaussian profile~\cite%
{Bwang2, Bwang}. Another assumes that the particles have no memory of their
diffusion directions, but may keep a positive correlation in the diffusion
distance of each step~\cite{Chubynsky}. Most of these studies have focused
on the diffusion of particles, except Ref.~\cite{li}, where the authors have
studied a one-dimensional toy model consisting of particles and mobile walls
arranged alternatively in a line. Kicking a particle in the middle of the
system, the energy of the kick was found to spread in a non-Gaussian normal
diffusion manner at certain parameters.

An important question arises: How general is this phenomenon? May it occur
in the energy diffusion process in finite-temperature lattices? In this
paper we show that non-Gaussian, normal energy diffusion may be common in
disordered systems and occurs with delocalization. The Hamiltonian of the
model we adopt is $H=\sum_{k}h_{k}$ with~\cite{DharPRL} 
\begin{equation}
h_{k}={\frac{p_{k}^{2}}{2m_{k}}}+\frac{(x_{k}-x_{k-1})^{2}}{2}+\frac{%
cx_{k}^{2}}{2}+\frac{\nu (x_{k}-x_{k-1})^{4}}{4},
\end{equation}%
where $p_{k}$, $m_{k}$, and $x_{k}$ are, respectively, the momentum, the
mass, and the displacement from the equilibrium position of the $k$th
particle. The lattice constant is set to unity, so that the number of the
particles, denoted by $N$, also represents the length of the lattice. The
masses of the particles distribute randomly and uniformly in the interval of 
$(0.8,1.2)$. The parameter $\nu $ controls the nonlinear interaction
strength between particles; When $\nu $ vanishes, the model reduces to the
disordered pinned harmonic (DPH) model, in which all the normal modes are
localized if $c$ is large enough. Recently, Dhar and Lebowitz have studied
the heat conduction properties of this model by nonequilibrium simulations~%
\cite{DharPRL}. They found that the lattice shows a normal heat conduction
behavior even if the nonlinear interaction is weak, implying that nonlinear
interaction may induce a normal energy diffusive process. Conventionally,
one may assume that this normal energy diffusive process is Gaussian, but we
will show in the following it is non-Gaussian and is closely related to
delocalization.

\begin{figure*}[!]
\centering
\center\includegraphics[scale=0.62]{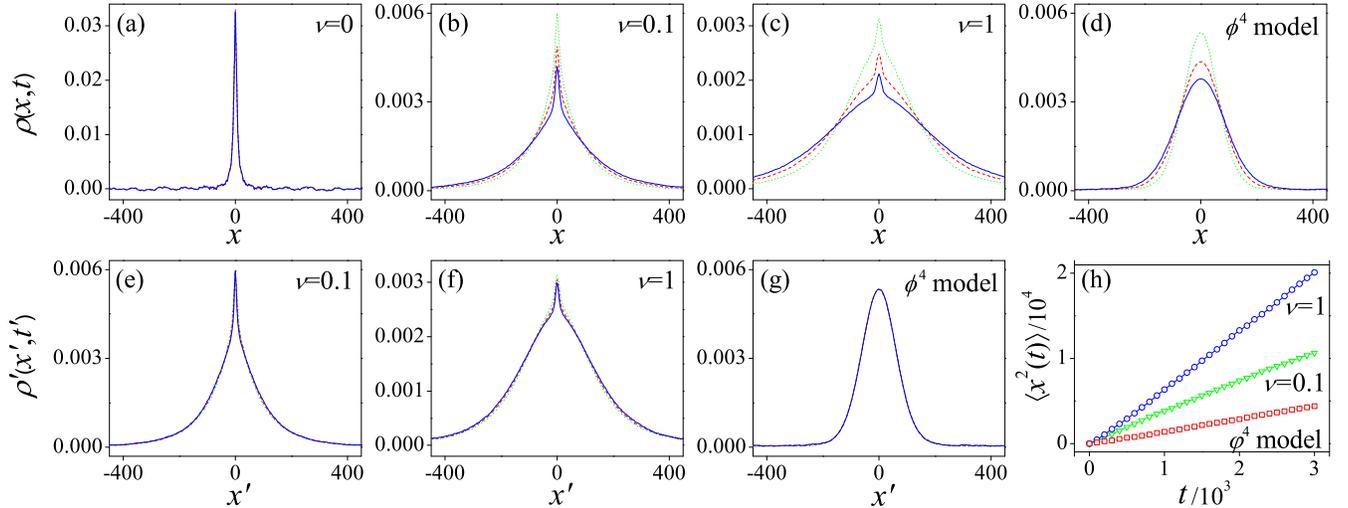}
\caption{(Color online) The spatiotemporal correlation functions of energy
density fluctuations at time $t=1000$ (green dotted line), $1500$ (red
dashed line), and $2000$ (blue solid line) for the disordered lattice model
with $\protect\nu =0$ (a), $\protect\nu =0.1$ (b), $\protect\nu =1$ (c), and
for the homogeneous $\protect\phi ^{4}$ model (d). Panels (e), (f) and (g)
present the rescaled results given in (b), (c) and (d), respectively, with $%
t_0=1000$ (see text), and panel (h) is for the mean squared displacement of
both models.}
\label{diffusion}
\end{figure*}

To start, we first determine the value of the parameter $c$ that guarantees
complete localization of normal modes in the DPH model. For this purpose, we
employ the Thouless criterion which is widely adopted to identify localized
and extended normal modes in condensed matter physics~\cite{thouless,
shengpng, somoza}. By simply changing the boundary condition from symmetric
(with $x_{1}=x_{N+1}$) to antisymmetric (with $x_{1}=-x_{N+1}$) and
calculating the frequency change, $\Delta \omega$, of a normal mode, the
Thouless criterion states that it is localized if $\Delta \omega =0$ or
extended otherwise. In this way, we find that $c\geq 1$ can guarantee
complete localization with the localization length being smaller than $3000$%
. As such in the following we will focus on the case of $c=1$ and $N=4096$.
In addition, the energy density is set to be $1.5$ as in Ref.~\cite{DharPRL}.

To explore the diffusion behavior of energy in the equilibrium state at a
non-zero, finite temperature, we investigate the spatiotemporal correlation
function of energy density fluctuations. First of all, by numerically
integrating the canonical motion equations, we evolve the system from a
randomly assigned initial condition for a sufficient long time ($>10^{6}$)
to ensure that it has relaxed to the equilibrium state. We then divide the
lattice into $N_{b}$ bins of width $b$ (in our simulations it has been set
to $b=4$) and calculate the spatiotemporal correlation function~\cite{Zhao,
ShundaE} 
\begin{equation}
\rho (x,t)=\frac{\langle \Delta e_{j}(t)\Delta e_{i}(0)\rangle }{\langle
\Delta e_{i}(0)\Delta e_{i}(0)\rangle }+\frac{1}{N_{b}-1},
\end{equation}%
where $x\equiv (j-i)b$ is the distance between the $i$th and the $j$th bin
and $\Delta e_{j}(t)=e_{j}(t)-e$ represents the energy fluctuation in the $j$%
th bin. $e_{j}(t)\equiv \sum h_{k}$, the summation involves all the
particles in the $j$th bin at time $t$ and $e$ is the average energy of a
bin. Because of the conservation of the total energy, there is a non-causal
correlation term $-1/(N_{b}-1)$ between different bins which should be
excluded. This consideration leads to the last term in the equation~\cite%
{ShundaE}. The function $\rho (x,t)$ thus gives the causal relation between
a local fluctuation and the effects it induces at another position at a
later time, hence it is in essence equivalent to the probability density
function that describes the diffusion process of the fluctuation. Taking
advantage of this equivalence, $\rho (x,t)$ has been successfully applied to
detecting the energy relaxation~\cite{Zhao, ShundaE} as well as the
localization properties of a system at a non-zero, finite temperature~\cite%
{wangPRE}.

Figure~1(a) shows $\rho (x,t)$ for the DPH model (with $\nu=0$) at different
times. It can be seen that $\rho (x,t)$ does not decay, which is a clear
signal of the energy localization. Meanwhile, it also indicates that $%
\rho(x,t)$ is an effective tool for identifying the localization feature in
a non-zero temperature environment.

\begin{figure}[!]
\centering
\includegraphics[scale=0.32]{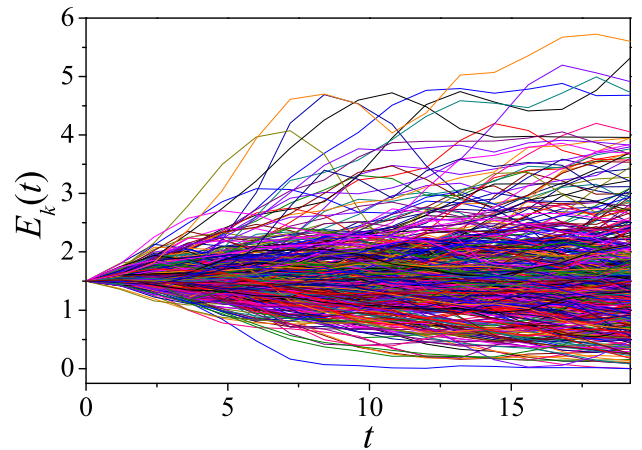}
\caption{(Color online) The energy of the normal modes of the disordered
lattice with the nonlinear interaction parameter of $\protect\nu =0.1$. The
diffusion behavior is a signal of delocalization.}
\label{spa}
\end{figure}

\begin{figure*}[tbp]
\centering
\center\includegraphics[scale=0.62]{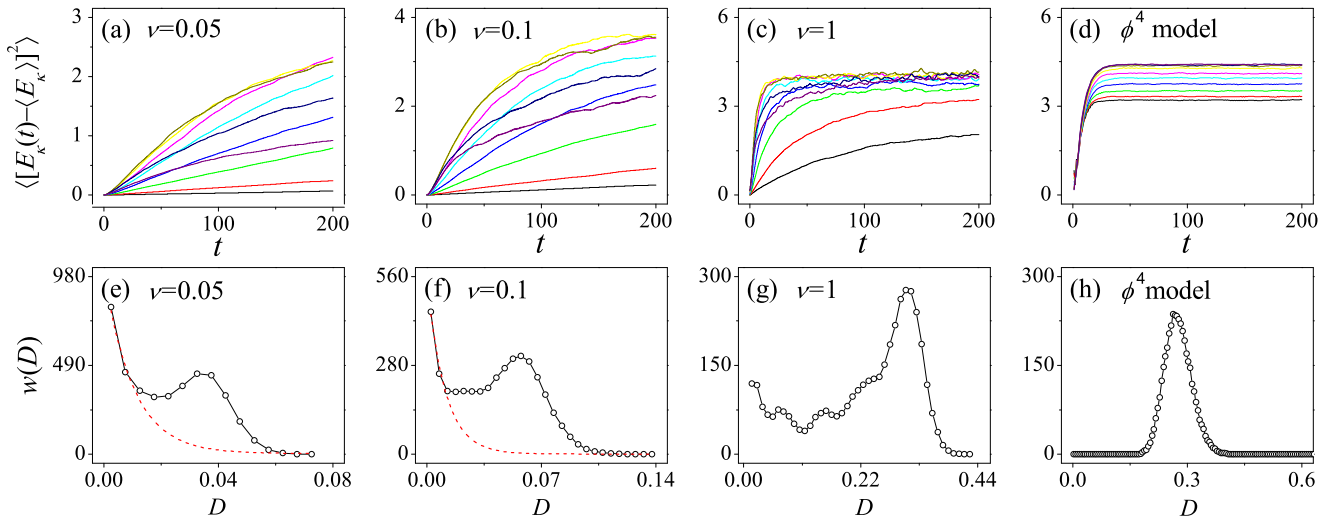}
\caption{(Color online) The energy diffusion among normal modes. Panel
(a)-(c) show $\langle [E_{k}(t)-\langle E_{k}\rangle]^{2}\rangle$ for the
disordered lattice model with $\protect\nu =0.05$, $0.1$ and $1$,
respectively, and (d) for the homogeneous $\protect\phi ^{4}$ model. Modes
of $k=409i$ with $i=1,...,10$ are taken as examples. Panel (e)-(h) show the
distribution of the diffusion coefficient corresponding to (a)-(d). The
dashed line in (e) and (f) indicates $\sim \exp(-aD)$, which is plotted for
reference.}
\label{diffusion}
\end{figure*}

Introducing nonlinear interaction into our model may cause delocalization.
Fig.~1(b) and (c) show $\rho (x,t)$ with $\nu =0.1$ and $1$, respectively.
It can be seen that delocalization does occur and as a consequence, the
height of $\rho (x,t)$ decreases as time (while the area under the $\rho
(x,t)$ curve keeps unchange since the energy is conserved). For the sake of
comparison, Fig.~1(d) shows $\rho (x,t)$ of a homogeneous lattice, i.e., the 
$\phi ^{4}$ model, with $h_{k}={\frac{p_{k}^{2}}{2}}+\frac{%
(x_{k}-x_{k-1})^{2}}{2}+\frac{x_{k}^{2}}{2}+\frac{x_{k}^{4}}{2}$. The $\phi
^{4}$ model has been shown to obey the Fourier heat conduction law~\cite%
{HuLiZhao}.

We find that the function $\rho (x,t)$ of both systems follow the same
scaling. By letting $t\rightarrow t^{\prime }=t$, $x\rightarrow x^{\prime
}=x/\xi $, and $\rho \rightarrow \rho ^{\prime }=\xi \rho $ with $\xi =\sqrt{%
t/t_{0}}$ ($t_{0}$ is a reference time), the curves of $\rho ^{\prime
}(x^{\prime },t^{\prime })$ at different times overlap perfectly [see
Fig.~1(e)-(g)]. This scaling implies that $\langle x^{2}\rangle =\int
x^{2}\rho (x,t)dx=(t/t_{0})\langle x^{2}(t_{0})\rangle $, i.e., the
dependence of the MSD on time is linear. As Fig.~1(h) shows, the linear
time-dependent behavior of the MSD can also be confirmed by calculating the
MSD directly. Based on the results of the MSD, we know the energy diffusion
is normal in both models. Meanwhile, in the homogeneous $\phi ^{4}$ model, $%
\rho (x,t)$ appears as a perfect Gaussian function, and therefore the energy
diffusion in this model can be identified as the normal diffusion in the
conventional sense.

However, $\rho (x,t)$ in the disordered model is obviously non-Gaussian. To
explore the reason, we study the relaxation behavior of localized modes
under nonlinear interactions. We first evolve the system up to a sufficient
long time to have it relaxed to the equilibrium state, then we project the
state of the system onto the normal modes of the DPH model, and record the
time series of energy for each mode, $E_{k}(t)$, $k=1,...,N$. With the
vanishing nonlinear parameter, i.e., in the case of the DPH model, $E_{k}(t)$
is a time-independent constant governed by the initial condition as a
consequence of localization. But with nonlinear interaction, energy exchange
among different modes should take place and result in the diffusive motion
of $E_{k}(t)$ in the energy space. As an example, in Fig.~2 we show $%
E_{k}(t) $ with $k=8i$ for $i=1,...,500$ at $\nu =0.1$. In this plot, the
starting time for each $E_{k}(t)$ curve is set to be the time when $E_{k}(t)$
is sufficiently close to $\left\langle E_{k}\right\rangle $ with $%
|E_{k}(t)-\left\langle E_{k}\right\rangle |/\left\langle E_{k}\right\rangle
<0.01$ for the first time, where $\left\langle E_{k}\right\rangle$
represents the average energy of the $k$th mode at equilibrium. This
treatment guarantees that the subsequent diffusion of $E_{k}(t)$ is around
this reference energy. In the perfect equilibrium state with the energy
equipartition, $\left\langle E_{k}\right\rangle $ should be the same for all
the modes~\cite{note2}. Fig.~2 indicates that delocalization takes place for
all the modes. The diffusion is finally bounded since the total energy of
the system is finite.

To further explore the diffusion rates of the normal modes we study the MSD $%
\langle \lbrack E_{k}(t)-\langle E_{k}\rangle ]^{2}\rangle $ as a function
of time. Fig.~3 shows the MSD for $k=409i$ with $i=1,...,10$ as examples.
Fig.~3(a)-(c) are for $\nu =0.05$, $0.1$ and $\nu =1$, respectively, and
Fig.~3(d) shows the MSD for the homogeneous $\phi ^{4}$ model. The MSD
curves are bounded finally as the total energy of the system is conserved.
The fact that the MSD curves do not converge to the same value in the $\phi
^{4}$ model suggests that energy equipartition is not well fulfilled in this
model. This phenomenon has been evidenced in this model when modes with
smaller $k$ have lower average energy at the equilibrium state~\cite{Pettini}%
. In the case of the disordered lattice model, by tracking a sufficient long
time we have found that $E_{k}(t)$ curves do mix with each other. In a
shorter period of $t\leq 10$, $\langle \lbrack E_{k}(t)-\langle E_{k}\rangle
]^{2}\rangle $ depends on time almost linearly. We fit the slop in this
period to measure the diffusion coefficient (denoted by $D_{k}$ for the $k$%
th mode). In general, the values of $D_{k}$ for different modes are
different. Fig.~3(e)-(g) show the distribution of the diffusion coefficient
[denoted by $w(D)$]; It can be seen that $w(D)$ approaches an exponential
decaying manner of $\sim A\exp (-aD)$ when $\nu \rightarrow 0$ in the two
extreme ranges of $D$, meanwhile, there is a Gaussian-like peak in the range
of moderate $D$. As $\nu $ increases, the peak becomes more and more
dominant. In the case of the homogeneous $\phi ^{4}$ model, $w(D)$ shows
only a Gaussian peak.

As mentioned previously, in order to explain the non-Gaussian normal
diffusion phenomenon observed in particle motion, some authors have assumed
that a particles experience different diffusion processes with different
diffusion coefficients and each individual process follows the normal
Brownian motion~\cite{Bwang2, Bwang}. For our disordered lattice model a
similar mechanism works. In fact, an initial energy fluctuation in the $i$th
bin can be decomposed into normal modes, i.e., $\Delta e_{i}(0)\sim
\sum_{k=1}^{N}\Delta e_{i}^{(k)}(0)$, where $e_{i}^{(k)}(0)$ denotes the
component of the $k$th mode. Indeed, as shown in Fig.~3, these components
proceed the normal diffusion with different diffusion coefficients, but it
is in the energy space rather than along the lattice. So we assume that a
component proceeds the normal diffusion in $x$ space as well, i.e., $\rho
^{(k)}(x,t)=(4\pi \overline{D}_{k}t)^{-\frac{1}{2}}\exp (-\frac{x^{2}}{4%
\overline{D}_{k}t})$, and its diffusion coefficient, $\overline{D}_{k}$, is
proportional to $D_{k}$ as $\overline{D}_{k}=\eta D_{k}$ with $\eta $ being
a constant. With this assumption, the PDF of $\Delta e_{i}(0)$ at time $t$
is given by $\rho (x,t)=\frac{1}{N}\sum_{k=1}^{N}\rho ^{(k)}(x,t)$, and $%
\langle x^{2}\rangle =\frac{\eta }{N}(\sum_{k=1}^{N}D_{k})t$. In other word,
if only each component proceeds the normal diffusion, the MSD will be a
linear function of time.

\begin{figure}[tbp]
\centering
\includegraphics[scale=0.31]{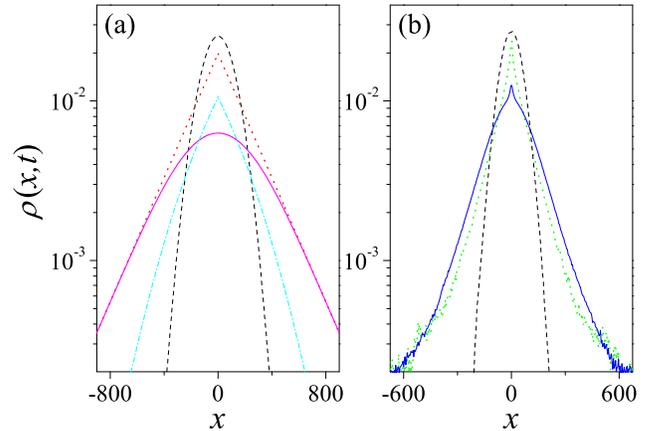}
\caption{(Color online) Panel (a) shows the profile of $\protect\rho (x,t)$
based on Eq.~(3) respectively for $w(D)$ being Gaussian (dashed line), for $%
W(D)\sim \exp (-aD)$ (dotted line), for $w(D)=0$ if $D<D_{0}$ and $w(D)\sim
\exp (-D/D_{0})$ if $D\geq D_{0}$ (solid line), and for $w(D)$=constant if $%
D<D_{0}$ and $w(D)=0$ if $D\geq D_{0}$ (dash-dotted line). In panel (b), $%
\protect\rho (x,t)$ in Fig.~1(b)-(d) with $t=1500$ is plotted in the
semi-log scale with the dotted, solid, and the dashed line, respectively. }
\label{spa}
\end{figure}

With this assumption we can also relate the profile of $\rho (x,t)$ to $w(D)$%
. Replacing the sum by integral in the thermodynamical limit of $%
N\rightarrow \infty $, we have 
\begin{equation}
\rho (x,t)=\int_{0}^{\infty }w(D)\rho ^{(k)}(x,t)dD.
\end{equation}%
In Fig.~4(a), $\rho (x,t)$ obtained with this formula for some assumed $w(D)$
has been plotted. For $w(D)$ being a Gaussian function, as in the case of
the homogeneous $\phi ^{4}$ model, $\rho (x,t)$ would be a Gaussian
function. For $w(D)\sim \exp (-aD)$, we have $\rho (x,t)\sim t^{-\frac{1}{2}%
}\exp (-\frac{|x|}{Ft})$ ($F$ is a constant) and $\rho (x,t)$ has a $\Lambda 
$ shape~\cite{Bwang}. More importantly, if only $w(D)$ has a constant
density close to $D=0$, the PDF would show a non-Gaussian profile. Fig.~4(a)
gives such an example with $w(D)$=constant for $D<D_{0}$ and $w(D)=0$ for $%
D\geq D_{0}$. We can see that $\rho (x,t)$ has a similar profile as the case
of $w(D)\sim \exp (-aD)$. The reason is that components with close-to-zero
diffusion coefficients relax slowly, and the smaller is $D_{k}$, the slower
the relaxation is. On the contrary, if there are no components whose
diffusion coefficients are close to zero, the center of $\rho (x,t)$ will
show a Gaussian shape. In Fig.~4(a) $\rho (x,t)$ for $w(D)=0$ for $D<D_{0}$
and $w(D)\sim \exp (-aD)$ for $D\geq D_{0}$ has been shown as an example;
while the tails of $\rho (x,t)$ are exponential, the center is Gaussian.

Therefore, components with close-to-zero diffusion coefficients are crucial
for the non-Gaussian profile, and delocalization may provide such
components. In our disordered model, when the nonlinear interaction is
suppressed, all modes are localized and the corresponding diffusion
coefficients are zero. But when the nonlinear interaction is activated,
delocalization happens but could be sufficiently slow if the nonlinear
interaction is weak enough. As $w(D)$ is not zero near the origin even for $%
\nu =1$, it explains why $\rho (x,t)$ shows a sharp peak around the origin
[see also Fig.~4(b)]. In more detail, when nonlinear interaction is weak,
the two ends of $w(D)$ fit the exponential decay [see Fig.~3(e)-(f)] and $%
\rho (x,t)$ has a $\Lambda $ shape, suggesting that the close-to-zero
diffusion components play the key role. When nonlinear interaction is
stronger, the center peak of $w(D)$ becomes remarkable, and as a
consequence, $\rho (x,t)$ deviates from the $\Lambda $ shape.

In summary, relaxation of local energy fluctuations in the studied
one-dimensional disordered system is featured by non-Gaussian normal
diffusion. This finding suggests that non-Gaussian normal diffusion may be
observed not only in certain specific complex systems but also in lattice
systems, and not only in particle motion but also in energy relaxation. The
mechanism is associated with delocalization of the localized normal modes.
For weak enough nonlinear interaction, sufficient modes may relax
sufficiently slow, which is a necessary condition for the non-Gaussian PDF.

Non-Gaussian normal diffusion observed in particle motion is believed to be
a transient process in a proper time scale as diffusion is in principle a
Markov process~\cite{Bwang}. This conjecture may not apply to a lattice,
because the number of modes is proportional to the lattice size and thus can
be large enough, and small coefficients can be close to zero enough as well
if nonlinear interaction is sufficiently weak. Therefore, the non-Gaussian
PDF of energy can sustain long. Though a specific disordered lattice is
studied here, the revealed mechanism may play a role in more general
delocalization process.

This work is supported by the NSFC (Grants No. 11335006 and No. 10805036).


\begin{thebibliography}{99}
\bibitem{note2} One can instead study this problem by setting each mode the
same initial energy and observing the time series of $E_{k}(t)$. The results
are qualitatively the same but there is a long transient process towards the
natural equilibrium state.
\end{thebibliography}
\end{document}